\documentclass[10pt, final, conference, letterpaper]{IEEEtran}
\usepackage{times}
\usepackage[noadjust]{cite}
\usepackage{url}
\usepackage{multirow}
\usepackage{xspace}
\usepackage{tabularx}
\usepackage{epstopdf}
\usepackage{array}
\usepackage{courier}
\usepackage{color}
\usepackage{flushend}
\usepackage{graphicx}
\usepackage{float}
\usepackage{algorithm}
\usepackage{algorithmic}

\usepackage{enumitem}
\usepackage{etoolbox}

\setlist[itemize]{noitemsep}
\newcolumntype{$}{>{\global\let\currentrowstyle\relax}}
\newcolumntype{^}{>{\currentrowstyle}}

\long\def\com#1{}

\newcommand{\blind}[1]{}

\title{An Experimental Investigation of Hyperbolic Routing with a Smart Forwarding Plane in NDN}

\author{
\IEEEauthorblockN{Vince Lehman, Ashlesh Gawande}
\IEEEauthorblockA{University of Memphis\\\{vslehman, agawande\}@memphis.edu} \\

\IEEEauthorblockN{Beichuan Zhang}
\IEEEauthorblockA{The University of Arizona\\bzhang@arizona.edu}

\and
\IEEEauthorblockN{               }
\IEEEauthorblockA{          \\              \\} \\

\IEEEauthorblockN{Lixia Zhang}
\IEEEauthorblockA{UCLA\\lixia@cs.ucla.edu}\\

\and
\IEEEauthorblockN{Rodrigo Aldecoa, Dmitri Krioukov}
\IEEEauthorblockA{Northeastern University\\\{raldecoa, dima\}@neu.edu} \\

\IEEEauthorblockN{Lan Wang}
\IEEEauthorblockA{University of Memphis\\lanwang@memphis.edu} \\
}

\makeatletter
\patchcmd{\@maketitle}
  {\addvspace{0.5\baselineskip}\egroup}
  {\addvspace{-2.25\baselineskip}\egroup}
  {}
  {}
\makeatother

\begin{document}

\maketitle

\begin{abstract}
Routing in NDN networks must scale in terms of forwarding table size and routing protocol overhead.
Hyperbolic routing (HR) presents a potential solution to address the routing scalability problem, because it does not use traditional forwarding tables or exchange routing updates upon changes in network topologies.
Although HR has the drawbacks of producing sub-optimal routes or local minima for some destinations, these issues can be mitigated by NDN's intelligent data forwarding plane.
However, HR's viability still depends on both the quality of the routes HR provides and the overhead incurred at the forwarding plane due to HR's sub-optimal behavior.
We designed a new forwarding strategy called Adaptive Smoothed RTT-based Forwarding (ASF) to mitigate HR's sub-optimal path selection.
This paper describes our experimental investigation into the packet delivery delay and overhead under HR as compared with Named-Data Link
State Routing (NLSR), which calculates shortest paths.
We run emulation experiments using various topologies with different failure scenarios, probing intervals, and maximum number of next hops for a name prefix.
Our results show that HR's delay stretch has a median close to 1 and a 95th-percentile around or below 2, which does not grow with the network size.
HR's message overhead in dynamic topologies is nearly independent of the network size, while NLSR's overhead grows polynomially at least.  
These results suggest that HR offers a more scalable routing solution with little impact on the optimality of routing paths.

\end{abstract} 
\section{Introduction} \label{sec:intro}

Named Data Networking (NDN~\cite{ndn-ccr2014}) is a data-centric Internet architecture that allows users to retrieve data directly by their names.
Thus, a native NDN routing scheme needs to support forwarding using data names.
Due to the sheer number of content names in today's Internet, the number of FIB entries in an NDN network could be prohibitively large, and the routing update overhead required to maintain consistent FIBs of this size can also be costly.
One of the major challenges in realizing NDN is to bound the size of routing state while supporting an unbounded data namespace.

At the same time, the requirements for routing protocols in an NDN network also differ from that in an IP network~\cite{Cheng:RoleOfRouting}.
NDN's intelligent forwarding plane can quickly steer packets around failures and discover best paths through active probing~\cite{Forwarding-COMCOM}.
Consequently, \emph{fast routing convergence is no longer a requirement} in an NDN network, and \emph{the optimality of the paths selected by routing protocols is not as important as in an IP network}.
However, as shown by Yi et al.~\cite{Cheng:RoleOfRouting}, a routing protocol is still needed to help find working paths faster with lower cost as compared to blindly probing all possible paths.

This paper investigates the viability of applying hyperbolic routing (HR) to NDN networks.
HR is greedy geometric routing based
on hyperbolic coordinates of nodes that encode network geometry (\cite{GreedyHyper,BoPa10}).
Assuming each node knows its own and its neighbors' coordinates and each Interest carries coordinates corresponding to its data name prefix, a forwarder simply computes the neighbors' distance to the coordinates and forwards Interests to the best next-hop(s) as measured by hyperbolic distance.
The following explains why we consider HR a potential candidate to scale NDN routing.

First, because both the distance calculation and best next-hop selection in HR can be done in real-time on a per-packet basis, a node does not need to maintain a full FIB -- it can simply keep a small cache of calculated routes to avoid repeated computation.  In other words, HR does not need to know the network's topological connectivity to compute routes and it does not incur the memory overhead of a full-size FIB.

Second, hyperbolic coordinates of a network are expected to be stable for a long time (months or years), eliminating the need for dynamic routing updates caused by short-term topology changes.
In today's IP network, it has been shown that, upon small- to medium-scale changes in network connectivity, \emph{HR with backtracking} can find alternative paths to destinations \emph{without nodes changing their coordinates} (\cite{GreedyHyper,BoPa10,greedyforward}).
With a good forwarding strategy, NDN's forwarding plane can also find alternative paths quickly in various failure scenarios~\cite{Forwarding-COMCOM}, providing the backtracking capability to support HR.

Third, HR seems more suitable for NDN than geometric routing schemes that rely on other network geometries as
the name spaces in an NDN network will likely evolve to hierarchical tree-like organization.  This organization appears to be a ubiquitous and often emergent feature of large-scale deployments, presumably because it allows for natural and essentially optimal representation of large-scale information spaces~\cite{Lamping95Hyperbolic}.
Instead of relying on provider-assigned names to facilitate aggregation, the main idea behind HR in NDN is to utilize the name space and network structure to scale routing by employing greedy routing using the underlying tree-like geometry. Although the NDN namespace is much bigger than IP's, we believe that by using hierarchical names, much like the URLs used to name today's web content, NDN can achieve a similar exponential reduction of space and overhead.
We note that the network and namespace geometry does not have to be exactly a tree, or even close to a tree (as measured by the treewidth metric, for example). According to the mathematical results on coarse hyperbolic geometry~\cite{Gromov07-book}, even very approximate tree-like structures (Gromov-hyperbolic~\cite{Gromov07-book}) can be mapped to hyperbolic spaces with low distance distortion.

The scalability benefits of HR, however, come with a price.
First, HR does not guarantee to find best paths for all destinations.
Second, HR can suffer from \emph{local minima} where packets get stuck at a node with no neighbor closer to the destination than itself (this can happen with or without topological changes).
Both issues can in principle be handled by NDN's smart forwarding plane, but a thorough investigation is necessary in order to answer the following questions:
(a) can HR with adaptive forwarding provide performance that closely approximates that of a shortest path routing protocol as measured by packet delay?  and
(b) is the cost of using HR, i.e., the forwarding plane overhead due to active probing,
much lower than the overhead under conventional routing protocols?

To answer these questions, we first developed a forwarding strategy called Adaptive Smoothed RTT-based Forwarding (ASF) that handles suboptimal routes and loops under HR.  ASF is a relatively simple strategy so as to give us a baseline performance of HR -- if such a basic design can lead to good performance, then any further improvement can only strengthen HR's viability for NDN.

We compared HR (in combination with ASF) with link state routing in an NDN network, assuming static hyperbolic coordinates and small-scale connectivity changes.
We measured the delay and loss rate experienced by NDN ping packets, as well as the probing overhead of ASF and update message overhead of link-state routing, in a 22-node NDN testbed topology and four scaled-down Internet topologies with 41, 58, 78 and 99 nodes.
We found that HR/ASF can \emph{provide paths that are comparable with those of link state routing}, achieving a similar packet delay and loss rate, even under frequent failure scenarios.   The delay stretch of HR has a median close to 1 and a 95th-percentile below or around 2.  More importantly, our experiments show that
\emph{HR's per-node message overhead in dynamic topologies grows very lowly as the network size increases}, while \emph{NLSR's per-node overhead grows at least polynomially}, suggesting that HR can scale to much larger topologies.

\section{Background}
\label{sec:background}

We examine hyperbolic routing performance by comparing it with NLSR (\cite{NLSR16}).
Here we provide some background information on NDN and the two routing approaches.

\subsection{NDN}

NDN enables users and applications to directly \textit{fetch data identified by a given name}.
A consumer puts the name of a desired piece of data into an
\textit{Interest} packet and sends it to the network. Routers use this name to forward the Interest toward the data producer(s).
Once the Interest reaches a node that has the requested data, the node
will return a \emph{Data} packet. This Data packet follows in reverse
the path taken by the Interest to get back to the requesting consumer.

To forward the Interest and Data packets, each
NDN router maintains three data structures:
a \emph{Forwarding Information Base (FIB)} containing precomputed routes
(or a \emph{Route Cache} of recently computed routes in the case of HR),
a \emph{Pending Interest Table (PIT)} containing all the Interests
that have been forwarded and are waiting for the returning data, and
a \emph{Content Store (CS)} containing previously received Data.
The router also has a \emph{Forwarding Strategy} module which takes input from the FIB and observes data fetching performance to determine whether and where to forward each Interest packet.  In Section~\ref{sec:HR-NDN}, we explain the NDN forwarding process when hyperbolic routing is employed.

\subsection{NLSR}

NLSR (\cite{NLSR16}) is a link-state routing protocol that uses NDN's Interest-Data packet exchanges to propagate reachability to name prefixes instead of IP prefixes. Its operations are similar to that of OSPF -- link-state advertisements (LSAs) are propagated throughout the entire network so that each router can build a complete network topology.  On the other hand, unlike OSPF, NLSR computes multiple (non-equal-cost) paths for each name prefix as an input to NDN's forwarding strategy.

NLSR propagates two types of LSAs -- Adjacency LSA and Prefix LSA. The Adjacency LSA is used to advertise all active links connecting one NDN router to its neighbors. The Prefix LSA, on the other hand, is used to advertise all the name prefixes that have been registered with the router.  The Link-State Database (LSDB) contains all the LSAs, and LSA dissemination is viewed as a data synchronization problem of the LSDBs maintained by the routers.  ChronoSync~\cite{chronosync} is used to synchronize the LSDBs on all the nodes.

\subsection{Hyperbolic Embedding}

The initial motivation for HR comes from the seminal observation in~\cite{kleinberg07infocom} that \emph{any} network topology can be embedded into the hyperbolic plane such that greedy geometric routing using the embedded node coordinates never gets stuck at local minima, i.e.\ all sources can reach all destinations if a forwarder selects as the next hop its neighbor closest to the destination in the hyperbolic plane.
In practical terms this means that a forwarder does not have to keep next-hop routes to all the destinations in the network in its FIB.
The only information required for forwarding decisions is the coordinate of the destination in the packet and the coordinates of the forwarder's neighbors, leading to exponential reduction in FIB space requirements. 
Unfortunately, upon a topology change, the embedding in~\cite{kleinberg07infocom} must be recomputed from scratch. 

To deal with this problem, several proposals have been investigated.
In~\cite{CvCr09}, re-embedding is optimized using back-pressure and other techniques.
An entirely different embedding strategy is proposed (\cite{BoPa10,PaPs12,PaPs14,PaKr15}) based on the existence of geometric similarities between hyperbolic spaces and large-scale Internet-like topologies; both exhibit hierarchical tree-like organization (\cite{geometry,PaBoKr11}).
The main benefit of this embedding strategy is that it exploits the intrinsic hyperbolic geometry of Internet-like networks~\cite{geometry} to find the congruent coordinates of nodes on the hyperbolic plane.  One outcome of this strategy, implemented using maximum-likelihood techniques (\cite{BoPa10,PaPs14}), is that the resulting embedding is remarkably robust.
Even if network connectivity changes significantly but the nodes do not change their coordinates, greedy forwarding using these static coordinates can still find, with high probability, alternative paths to destinations if they exist (\cite{GreedyHyper,BoPa10}).
In other words, even if the network topology is dynamic, the network embedding and node coordinates do not have to be recomputed, so HR does not incur the overhead of traditional routing protocols upon network topology changes.
The cost paid for this routing overhead reduction in dynamic networks is some slight increase of path stretch and decrease of success ratio, defined as the percentage of source-destination pairs reachable via greedy forwarding (\cite{GreedyHyper,BoPa10}).

\section{Hyperbolic Routing in NDN}
\label{sec:HR-NDN}

To provide context for our forwarding strategy and experiment design, below we describe how NDN forwarding may work under hyperbolic routing and how hyperbolic coordinates may be computed and distributed in NDN.

\subsection{Interest Forwarding Process}
\label{sec:interest-forward}

\begin{figure}[tbh]
\centering
\includegraphics[width=\columnwidth]{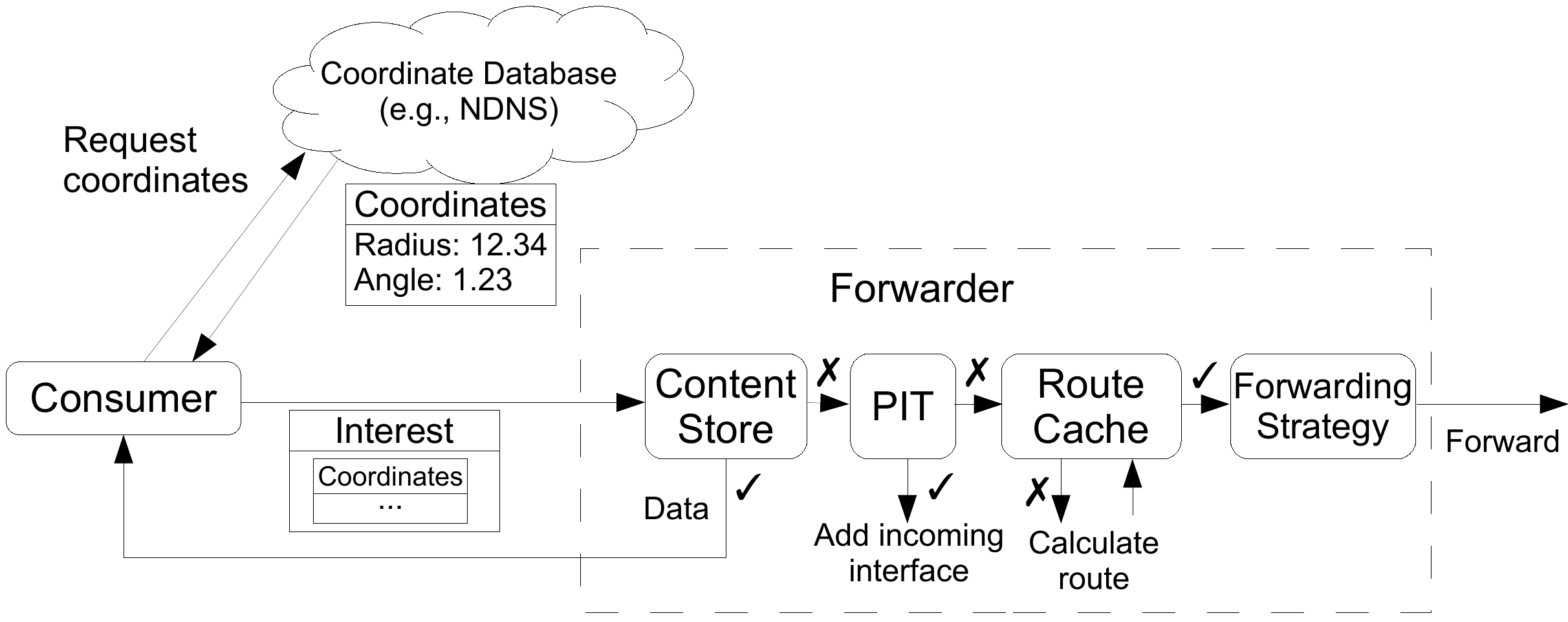}
\vspace{-6mm}
\caption{Interest Forwarding under Hyperbolic Routing}
\label{fig:hr-forwarding}
\vspace{-2.5mm}
\end{figure}

Figure~\ref{fig:hr-forwarding} shows how an Interest may be forwarded by a router when hyperbolic routing is employed.
Suppose the consumer wants to retrieve the data identified by a particular name, it first obtains the hyperbolic coordinates for the name, e.g., from a distributed database such as the NDNS~\cite{NDNS} (see Section~\ref{sec:hyper-corr}).
It then puts the data name and the coordinates in an Interest, and sends the Interest to its connected router.
When the router receives the Interest, it first checks the Content
Store for matching data; if a match is found, the router returns the Data packet
on the interface from which the Interest came.  Otherwise, the router
looks up the name in its PIT, and if a matching entry exists, it simply
records the incoming interface of this Interest in the PIT entry.
In the absence of a matching PIT entry, the router checks the \emph{Route Cache}
to see whether next hops for the coordinates carried in the Interest have already been calculated; if not, the router calculates the next hops to reach the coordinates using the greedy geometric routing algorithm, and installs them in the Route Cache.
Finally, the router uses the forwarding strategy to select one or multiple next hops to forward the Interest.

Because a large number of next hops may increase the size of the Route Cache and time for selecting a route, the network operator may specify an upper limit on the number of next hops per name prefix, called \emph{Multi-path Factor} in this paper.  We show in Section~\ref{sec:eval} that this is a key factor in forwarding plane performance when hyperbolic routing is used.

Note that HR in NDN is \emph{very different} from location-based routing in IP. First, NDN Interests carry both data names and the corresponding coordinates, with the latter simply acting as hints to direct the forwarding.  As such an Interest does not necessarily reach the producer; as soon as the Interest hits a router with the named content in its content store (the probability of which likely increases as the Interest gets closer to the producer), the content will be returned by that router. Second, HR in NDN is multi-path instead of single-path, i.e., each router calculates multiple routes to the coordinates carried in an Interest, so that if one greedy next hop fails to bring back the data, other next hops can be explored.

\subsection{Assigning and Disseminating Hyperbolic Coordinates}
\label{sec:hyper-corr}

The assignment of hyperbolic coordinates for the current Internet topology has been studied extensively (\cite{BoPa10,PaPs12,PaPs14,PaKr15}).
However, both the process and algorithm to assign hyperbolic coordinates to NDN names are still open questions and part of our ongoing work.
One difficulty is that there is currently no large-scale NDN deployment, or any reasonable model of it, to experiment with.   

For our experiments, we adopt the following approach to assign the coordinates.  First, since name prefixes are associated with their producer sites, to be able to hyperbolically route NDN Interests toward the producers, it is natural to map a name prefix to the hyperbolic coordinate of the producer site. In other words, hyperbolic routing to an NDN name prefix relies on (but is not equal to!) greedy forwarding toward the hyperbolic coordinate of its producer site.
Second, previous simulation work by Aldecoa and Krioukov~\cite{greedyforward} has shown that hyperbolic coordinates derived from the Internet AS-level topology using the HyperMap coordinate inference algorithm~\cite{PaPs14} works reasonably well on the NDN testbed topology -- each site is assigned the hyperbolic coordinates of the AS to which it belongs, and even with single-path greedy forwarding, the success ratio for packet delivery is $82\%$.
Given that the connection between the testbed topology and AS-level coordinates is rather indirect, the results are unexpectedly good, encouraging us to test HR using this set of coordinates with multi-path forwarding on the real NDN platform.

Our rationale for the above strawman approach is as follows.  If the AS-level coordinates indeed lead to satisfactory forwarding performance in our experiments under realistic NDN settings, any improvement in the coordinate assignment algorithm will result in even better performance, which justifies further study of hyperbolic routing in NDN.  In addition, we can immediately use this approach on the current NDN testbed to support longer-term evaluation of hyperbolic routing without waiting for a better coordinate assignment algorithm.

We expect hyperbolic coordinates to be computed offline based on local information before a new name prefix is added to the network and occasionally thereafter to adapt to long-term topological changes.  As shown in previous work (\cite{BoPa10,PaPs14}), a new node can compute its hyperbolic coordinate using only the coordinates of other nodes in the network. That is, quite surprisingly, the new node does not have to know anything about network topology other than what existing nodes it connects to. Furthermore, as shown in~\cite{PaKr15}, the accuracy of this coordinate computation is essentially indistinguishable from the same computation but using only the coordinates of few nodes that are hyperbolically closest to the new node. In other words, new nodes can compute their coordinates using only local information about network topology and geometry.

No routing information ever needs to be globally flooded through the network at any step of the above process.
New coordinates can be disseminated by mechanisms outside routing, e.g., using NDNS~\cite{NDNS}, a distributed database similar to DNS designed for NDN.
Upon computing new coordinates, a network operator can inject the name prefix and its coordinates into the network's own NDNS server.
The coordinates can be retrieved by any node on demand, similar to how DNS queries work.
Other options are possible, but their discussion is out of scope for this paper.

\section{Adaptive SRTT-based Forwarding} \label{sec:strategy}

Hyperbolic distance-based routing ranking may not reflect the actual forwarding delay for two reasons. First, 
it is possible that a next hop that is hyperbolically closer to the destination may actually lead to a longer path than one that is hyperbolically further from the destination. Second, when there are failures and recoveries in the network, HR does not adapt, thus its ranking may not reflect what is the shortest path at the moment. To have forwarding performance similar to that of shortest-path routing, we need a strategy that can find a shorter path if available and adapt to network changes.

The forwarding strategies implemented in the current NDN prototype focus on being able to retrieve contents, but may not find the shortest path in some situations.  For example, the Best-Route Strategy simply uses the next hop ranked highest by the routing protocol, which may not be optimal in the case of HR.
We compared the delay stretch in HR using the Best-Route Strategy against the delay in LS using the Best-Route strategy and confirmed that using the Best-Route strategy with HR results in both high delay stretch and a high loss rate (Section~\ref{sec:multpathfactor}).
As such, we developed a new strategy called Adaptive Smoothed RTT-based Forwarding (ASF) that chooses the best next hop based on RTT measurement, and also periodically probes other next hops to learn their RTTs.
This strategy is similar in spirit to the Red-Green-Yellow strategy proposed by Yi et al.~\cite{Forwarding-COMCOM}.
One important difference is that the latter strategy probes alternative paths only when routing ranking changes, which does not work with HR as HR's ranking does not adapt to short-term network changes. Another difference is that Yi's scheme does not consider measured RTT when choosing the best next hop, which may lead to suboptimal paths under HR.

While other proposed forwarding strategies (e.g., ~\cite{strategy-INFORM,probability-adaptive-forwarding, strategy-beyond-network-selection, strategy-ant-colony}) may be well suited for HR, they require modification of the Data packet~\cite{strategy-INFORM} or use more complex next hop selection processes, with multiple parameters, that can make analysis difficult~\cite{probability-adaptive-forwarding, strategy-beyond-network-selection, strategy-ant-colony}. 
We designed ASF to be easy to analyze, therefore we did not optimize its various parameters, such as probing period and probability, dynamically based on observed performance.  However, as we will show in Section~\ref{sec:eval}, it performs surprisingly well in our experiments.

\subsection{Best SRTT-based Forwarding}
\label{sub:best-srtt-forwarding}

\com {
\begin{algorithm}[tb]
\caption{Forwarding Interface Selection}
\label{alg:forward-face-select}
\footnotesize
\begin{algorithmic}[1]
\STATE $interfaces$ = $RouteEntry$.$nexthopInterfaces$
\STATE Initialize lists $first$, $second$, $third$
\FOR {each $face$ in $interfaces$}
	\IF {$face$.$hasSrtt$ and !$face$.$hasTimedOut$}
		\STATE Add $face$ to $first$ sorted by lowest $face$.$srtt$
	\ELSIF {!$face$.$hasSrtt$ and !$face$.$hasTimedOut$}
		\STATE Add $face$ to $second$ sorted by $face$.$cost$
	\ELSE
		\STATE Add $face$ to $third$ sorted by $face$.$cost$
	\ENDIF
\ENDFOR
\IF {!$first$.$isEmpty$}
	\STATE Use $first$.$front$ to forward Interest
\ELSIF {!$second$.$isEmpty$}
	\STATE Use $second$.$front$ to forward Interest
\ELSE
	\STATE Use $third$.$front$ to forward Interest
\ENDIF
\end{algorithmic}
\end{algorithm}
}
Every time a Data packet is received, we take a sample of the RTT, i.e., the time between when the Interest was sent and when the Data arrives. To accommodate fluctuations in RTT values, we compute the Smoothed RTT (SRTT), a moving average of the RTT samples, in the same way as TCP. 
We maintain one SRTT for each next hop in a route entry.

Given a route  entry, we divide its next hops into three groups.
Group 1 contains next hops that have SRTT values, i.e., they are working at the moment and returning Data. 
Group 2 contains next hops that do not have measurements yet. Group 3 contains next hops that are experiencing Interest timeouts.

When an Interest arrives, the strategy picks the next hop with the lowest SRTT in Group 1.
If this group is empty, the strategy will choose the next hop with the lowest routing cost from Group 2 or, if Group 2 is empty, from Group 3.
Group 2 is preferred over Group 3, because next hops that have never been used may work while next hops that are timing-out likely do not work.

\subsection{Probabilistic SRTT-based Probing}

As network conditions change over time, the shortest path may also change.
If an alternate next hop becomes better than the current next hop, we will not be able to discover it if we are not sampling its SRTT.
To solve this problem, ASF employs a periodic and probabilistic probing scheme to probe next hops not currently in use.
The first probe is scheduled to occur shortly after an entry is created in the Route Cache.  In our experiments, the delay is a random interval between [0, $T_1$] seconds, where $T_1$ is 5 by default.  This random delay prevents triggering probing at each router along the Interest's path at the same time.
After the first probe, the subsequent probes are scheduled to occur every $T_2$ seconds, which is set to 60 in our experiments (see Section~\ref{sec:eval}).

A probing Interest carries the same name as the original Interest but with a different nonce.
We use the same name for two reasons: a different name under the same prefix may be mapped to different coordinates (e.g., /google/sports and /google/music may have different coordinates);
and routers do not have a way to generate a valid new name -- if the generated name does not have associated data, then the router will mistakenly take the lack of Data as an indication that the path does not work.
We use a different nonce to ensure that if the probed path works, data will be returned on this path in addition to the primary path so that we can take an RTT sample of the probed path.
Note that Interest aggregation and caching may cause some RTT estimates to be inaccurate, but will not negatively affect the SRTT measurements in the long run,
because even if one RTT estimate returns an inaccurate RTT, subsequent measurements will cause the SRTT estimate to eventually converge to the correct one. Moreover, even if the strategy switches to an alternative path due to a false RTT measurement, eventually the good path will be probed and the strategy will switch back.

To quickly gain information about a next hop that has no measurements, ASF first chooses the next hop with the lowest routing cost from Group 2 to probe (see Section~\ref{sub:best-srtt-forwarding} about how the next hops are grouped). When Group 2 becomes empty, ASF will probabilistically choose a next hop from Group 1 and Group 3 to probe.
The next hops in Group 1 and 3 are ranked by SRTT and HR distance, respectively, while Group 1 has higher ranking than Group 3.
The probing probability of the next hop that has rank $i=1,\ldots,N$ in the sorted list (1 being the highest ranking) is
\begin{equation}
\label{formula:probing-probability}
P(i) = 2\frac{N + 1 - i}{N(N+1)}.
\end{equation}
In this way, next hops that performed better previously are more likely to be probed, thus allowing
the strategy to revert back to a better performing path after recovery from a failure.

\section{Experiment Methodology and Results}
\label{sec:eval}

In principle, data retrieval in NDN can operate without any routing since interfaces can simply be tried in sequence without any routing hints.
But, this strategy is equivalent to random walking through the network which would incur huge delays in large networks.
Shortest path routing minimizes delays but incurs the traditional high link-state routing overhead.
Thus, we focus our evaluation to study the delay-overhead trade-off space in comparison between NLSR and HR with the ASF strategy.   To study this trade-off space, we measure NLSR's routing overhead and HR/ASF's probing overhead, as well as the ping-traffic delay and loss, in packet-level emulation experiments in static and dynamic networks.
Since we are interested in routing \emph{scalability}, we need to evaluate how the delay and overhead numbers \emph{scale} with the growing network size~$N$. Therefore, we design experiments not only for a current NDN testbed snapshot ($N=22$), but also for a sequence of Internet-like topologies obtained from the current AS Internet topology by rescaling it down to different sizes with $N$ ranging from $N=41$ to $N=99$, the latter limited by our computational constraints. 

\subsection{Methodology}
\label{sec:method}

This work is intended to be the first step of our investigation and as such, we do not expect to draw a final conclusion on the feasibility of HR in NDN. 
Instead, we view the results from this study as an input to the next step.
Rather than looking at the combinations of all possible factors that may affect the performance of HR, we focus on a few so that we may gain an in-depth understanding of them.  Below, we describe and justify our experiment design.

\begin{figure}[tbh]
\centering
\includegraphics[width=\columnwidth]{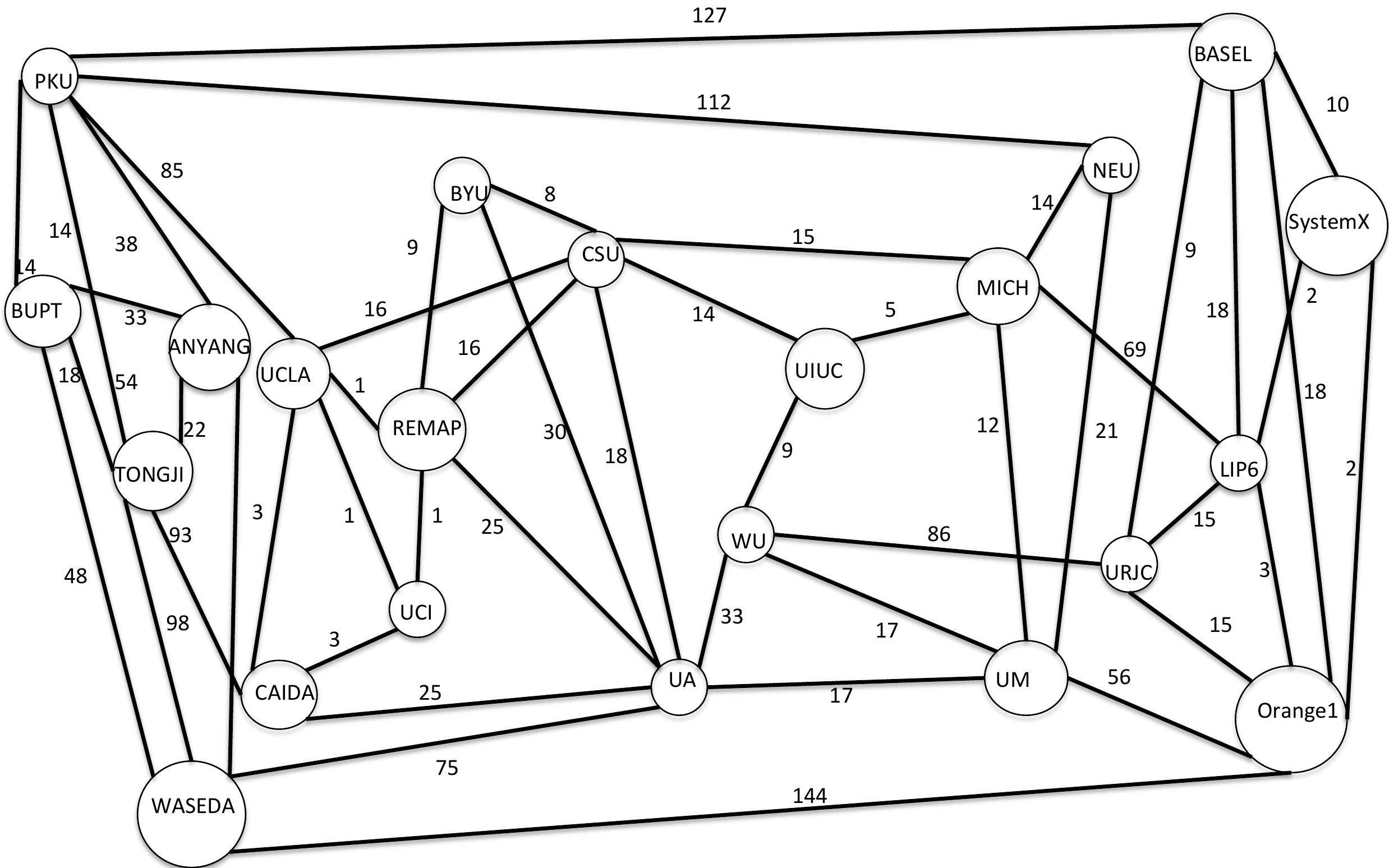}
\caption{Network topology with 22 nodes and 50 links based on NDN testbed (link cost is based on one-way delay between nodes)}
\vspace{-4mm}
\label{fig:topology}
\end{figure}

\medskip
\noindent\textbf{Topology, Network Size, and Traffic} -- one of our experiment topologies is a snapshot of the actual NDN testbed topology with 22 nodes and 50 links (Figure~\ref{fig:topology}).
The routing cost of each link is set to the delay between the two neighboring nodes. The hyperbolic coordinates of the nodes are set equal to the coordinates of the
AS to which these nodes belong~\cite{greedyforward}. The per AS coordinates are computed using the algorithms and data described in \cite{PaPs14} and \cite{PaKr15}.
Testbed nodes that belong to the same AS have small disturbances added to make the coordinates unique.
Each node advertises one name prefix and produces ping data under that name prefix. The \emph{ndnping} tool~\cite{ndn-tools-github} is used to send ping Interests from every node to every other name prefix once a second. 
All the ping names are unique, so caching will not affect the ping delays.

\begin{table}[bt]
\centering
\caption{Topologies derived from the Internet AS Topology}
\begin{tabular}{|c|c|c|c|c|}
\hline
\# Nodes (N) & 41 & 58 & 78 & 99 \\
\hline
\# Links (M) & 155 & 251 & 345 & 442 \\
\hline
Average Degree ($\bar{k}$) & 7.56 & 8.66 & 8.85 & 8.93\\
\hline
\end{tabular}
\vspace{-4ex}
\label{tab:topologies}
\end{table}

Furthermore, in order to test the scaling properties of HR, we used realistic Internet-like topologies of a varying size $N$ ranging between $41$ and $99$, the upper limit constrained by our computational resources. As shown in~\cite{SeKrBo08}, the AS Internet topology is self-similar, meaning that its subgraphs induced by the $N$ highest-degree nodes have all the structural properties of the original full topology, albeit renormalized to smaller values of $N$ and higher values of the average degree $\bar{k}$. That is, the smaller the size $N$ of these subgraphs, the larger their average degree $\bar{k}$~\cite{SeKrBo08}, an effect related to the rich-club phenomenon~\cite{Zhou2004RichClub}. 
For example, the subgraph consisting of $N=100$ highest-degree nodes has average degree $\bar{k}=40.78$, while the average degree in the full Internet topology ($N=37,618$) is $\bar{k}=5.22$. 
In the real Internet, however, the average degree is approximately constant as the Internet grows over years (\cite{BoPa10,PaPs14}).
Therefore, to construct Internet-like topologies of small varying size and approximately constant average degree, we start with the full Internet AS topology embedded in the hyperbolic plane using the HyperMap algorithm~\cite{PaPs14}. We extract its subgraphs induced by different numbers $N$ of highest degree nodes, and in each of these subgraphs, we then retain only $M=\bar{k}N/2$ ``shortest'' links, i.e., links between hyperbolically closest nodes, where $\bar{k}=5.22$ is the average degree in the full Internet topology. These shortest links are the most significant links in the topology, mostly between the pairs of nodes with the highest values of the product of their degrees~\cite{BoPa10}. Removing the rest of the links disconnects the subgraphs, so we extract the largest connected components from them. These largest components are then the networks on which we perform our routing scalability analysis. The values of $N$, $M$, and $\bar{k}$ are reported in Table~\ref{tab:topologies}.
To assign an appropriate delay to each link, the geographic coordinates between each AS associated with the nodes on a link is computed. The computed geographic distance is then used to approximate a delay for the link.

\medskip
\noindent\textbf{Hyperbolic Routing} -- For our evaluation purposes only, we have implemented HR within NLSR to distribute routers' hyperbolic coordinates. NLSR maps each name prefix to its originating router's
coordinates and then calculates a ranked list of next hops using the neighbor routers' coordinates. 
Note that in a real deployment, the coordinates are not flooded globally (see Section~\ref{sec:hyper-corr}).
To evaluate how well our proposed ASF strategy can help HR handle sub-optimal routes, we compared HR's performance under ASF with that under the
default Best-Route strategy which simply uses the best route selected by the routing protocol.  Another important variable we examine is how the multi-path factor affects forwarding performance when HR is used.  When more next hops are available in a route entry, the forwarding strategy may eventually find a next hop that is able to return Data.  
In our experiments, we evaluate HR using a multi-path factor of 2, 3, 4, and all possible next hops per prefix. 

\medskip
\noindent\textbf{Link State Routing Protocol} -- While we could have used another routing protocol to compare HR against, we chose link-state (LS) as it is simple to analyze and calculates shortest paths.
Since LS routing provides optimal paths, probing for alternative paths is unnecessary.
Thus, we can use the Best-Route forwarding strategy which always uses the highest routing ranked next hop to forward an Interest.
Indeed, we have verified in our experiments that, under LS, using ASF and Best-Route strategies provide the same performance.
Furthermore, with the Best-Route strategy, LS performs identically amongst tests with two, three, or all possible next hops per prefix, as only the highest ranked next hop is used.
We therefore use NLSR with the Best-Route strategy and a multi-path factor of 2 in all of our comparisons with HR.

\medskip
\noindent\textbf{Environment} -- All our experiments were performed using Mini-NDN~\cite{minindnsite}, an NDN network emulation tool. 
In Mini-NDN, an entire experiment runs on a single machine, and each node in the network topology is executed in a container with its own resources.  

\medskip
\noindent\textbf{Metrics} -- we use the following three metrics for comparing packet delay and overhead:\\
\textbullet\ \textit{Delay Stretch} -  Delay stretch indicates the ratio between a packets delay in HR vs. its delay in LS.
A stretch of 1 indicates that HR is not incurring any additional delay over what is required in LS.
To calculate the delay stretch, the RTT of a ping in HR is divided by the RTT of the corresponding ping in LS for each node pair.
Then, for each second of the experiment, we calculate the median, 75th percentile, and 95th percentile of the delay stretches among all the node pairs.\\
\textbullet\ \textit{Loss Rate} - For each node, the loss rate is the total number of ping timeouts from that node to all other nodes in the network divided by the total number of pings
from that node.\\ 
\textbullet\ \textit{Message Overhead} - The total message overhead for link-state routing is calculated by adding together the SYNC Interests/Data and LSA Interests/Data
sent during the experiment.  Note that we ignore the NLSR message overhead during the initial routing convergence.   For HR, the total message overhead is 
equal to the number of probe Interests sent by ASF.
We do not include the overhead for distributing hyperbolic coordinates for two reasons.  First, these coordinates remain stable for months or years, and any messages triggered by their changes occur at time scales much larger than those of LSA and SYNC messages.  Second, we expect that these coordinates are retrieved on demand by consumers, and they can be cached by the consumers and any intermediate routers for a long time (given that they seldom change).

We also considered comparing the total number of forwarded Ping Interests and Ping Data packets under the two approaches to assess the cost of suboptimal paths in HR. However, this metric can be misleading as HR with shorter unsuccessful paths may appear to have a lower number of ping packets than NLSR with longer successful paths.

\subsection{ASF vs. Best-Route and The Impact of Multi-path Factor}
\label{sec:multpathfactor}

\begin{figure}
\centering
    \begin{minipage}{0.4\textwidth}
        \centering
        \includegraphics[width=\columnwidth]{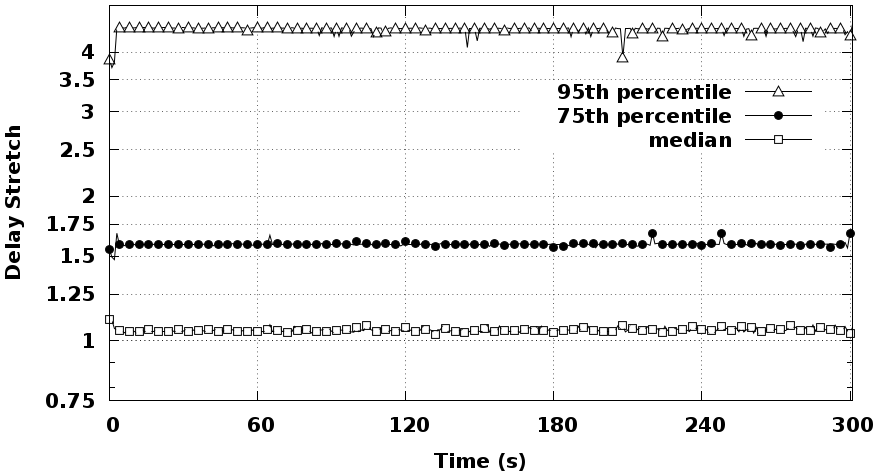}
        \mbox{\footnotesize a) HR with Best-Route Strategy}
        \vspace{2ex}
    \end{minipage}
    \begin{minipage}{0.4\textwidth}
        \centering
        \includegraphics[width=\linewidth]{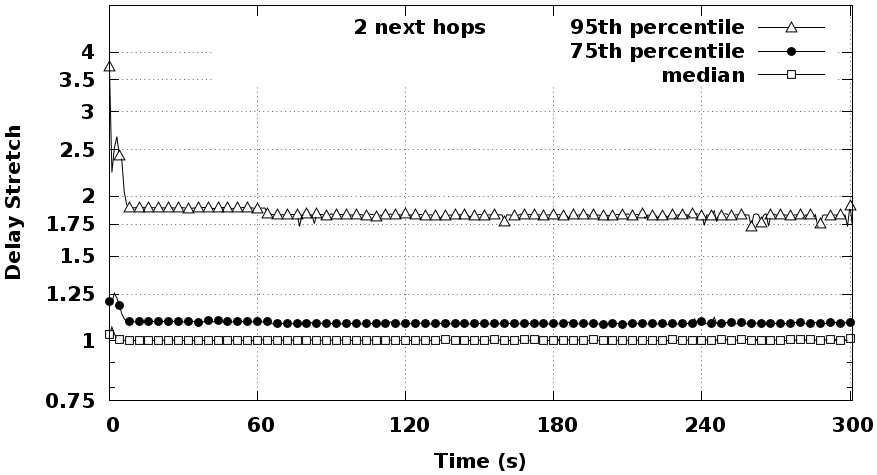}
        \mbox{\footnotesize b) HR with ASF and 2 next hops}
        \vspace{2ex}
    \end{minipage}
     \begin{minipage}{0.4\textwidth}
        \centering
        \includegraphics[width=\linewidth]{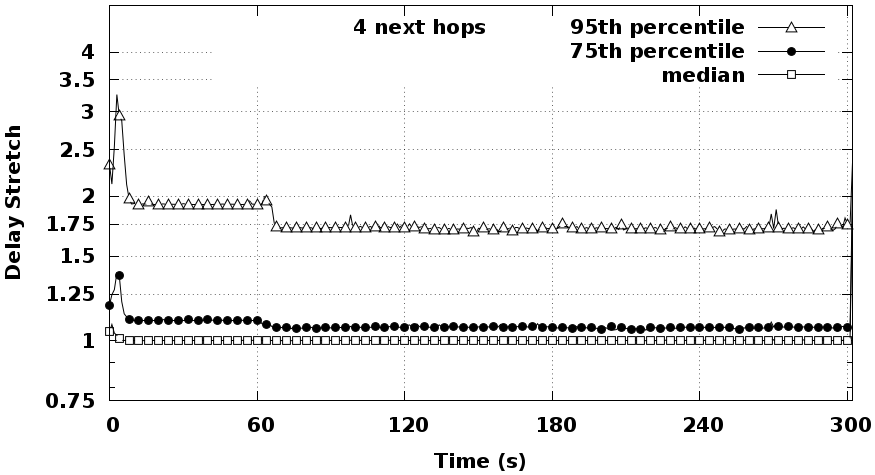}
        \mbox{\footnotesize c) HR with ASF and 4 next hops}
        \vspace{2ex}
    \end{minipage}
     \begin{minipage}{0.4\textwidth}
        \centering
        \includegraphics[width=\linewidth]{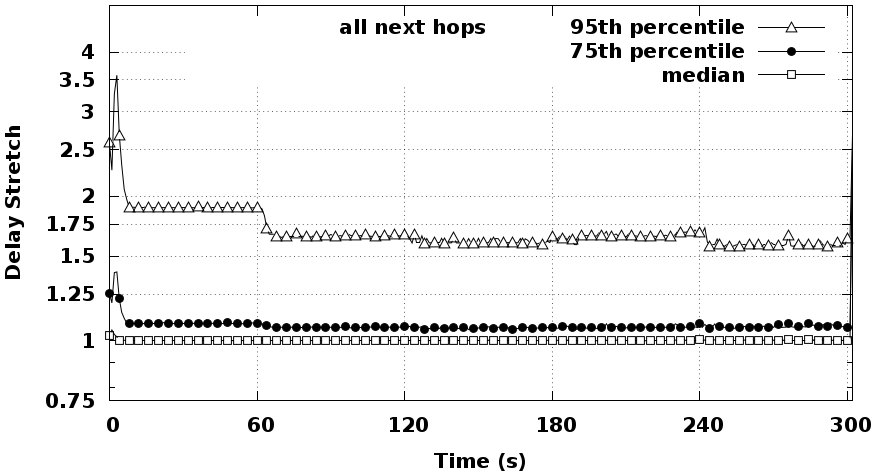}
        \mbox{\footnotesize d) HR with ASF and all next hops}
    \end{minipage}
    \caption{Delay stretch of HR compared to link-state routing with Best-Route strategy on 22-node testbed topology without failures.}
    \label{fig:no-failure-stretch}
   \vspace{-4ex}
\end{figure}

We first compare the ASF strategy with Best-Route strategy and then study the impact of multi-path factor.  
We use the 22-node testbed topology and let every node ping every other node once per second.  We use four different multi-path factors (2, 3, 4 and all) for forwarding and a probing interval of 60 seconds for ASF.  
The results for a multi-path factor of 3 are similar and therefore omitted for brevity.

Figure~\ref{fig:no-failure-stretch} shows the delay stretch of HR over link state\footnote{Note that we remove those pings that experienced timeouts from the calculation.  As such, the stretch curve may increase in the first few seconds as some ping pairs initially experienced timeouts in HR but through ASF probing the subsequent pings were able to find working paths.  When their stretch values were included in the calculation, there may be an increase in the various stretch statistics.}.  First of all, \emph{compared with the Best-Route strategy, ASF allows hyperbolic routing to achieve much lower delay stretch}.  Second, we make the following observations for HR with ASF (Fig~\ref{fig:no-failure-stretch}(b)-(d)): (1) the median stretch is almost 1 and the 75th-percentile stretch is slightly higher than 1. This shows that \emph{the packet delays under HR/ASF are quite close to those under link-state routing except in a small fraction of the cases}; (2) the 95th-percentile curve stabilizes after one probing interval of 60 seconds, meaning that the ASF strategy is quick in finding better paths when the initial ranking based on hyperbolic coordinates is suboptimal; (3) a higher multi-path factor leads to a slight increase in performance as the strategy has more choices to get out of loops and suboptimal paths, but \emph{a multi-path factor of 4 is nearly as good as all next hops}.

\begin{figure}
\centering
    \includegraphics[width=0.85\linewidth]{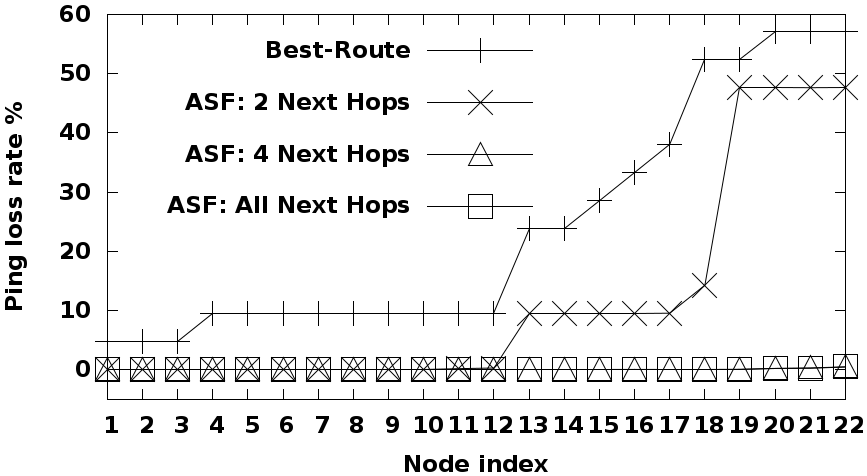}
    \vspace{-2ex}
    \caption{Loss Rate of HR on 22-Node Topology without Failures}
    \label{fig:no-failure-loss-rate}
\end{figure}

\begin{table}
\vspace{2ex}
\caption{Per-Node Message Overhead in 22-Node NDN Testbed Topology (Packets/Second)}
\begin{tabular}{|c||c|c|c|c||c|c|}
\hline
\multirow{2}{*}{Experiment} & \multicolumn{4}{c||}{No Failure} & \multicolumn{2}{c|}{Sequential Failure} \\
& LS  & \multicolumn{3}{c||}{HR} & LS & HR \\
\hline
\# Next Hops & 2 & 2 & 4 & all & 2 & 4 \\
\hline
Overhead & 1.05 & 1.01 & 0.98 & 0.96 & 2.80 & 0.33 \\
\hline
\end{tabular}
\vspace{-4ex}
\label{tab:ndn-testbed-overhead}
\end{table}

Although this experiment does not have any topological failures, losses may still occur in HR due to the multi-path factor, which imposes a limit on the number of next hops installed for each route entry.  If every next hop in a route entry causes a loop in the network, the strategy will not be able to find a next hop to get out of the loop. Note that even when all next hops are available to the strategy, the forwarding plane may still experience \emph{initial} losses while the strategy determines working paths.
Figure~\ref{fig:no-failure-loss-rate} shows the loss rate for each node under HR (there are no losses in link-state routing).
First, even with only 2 next hops, ASF has lower loss rate than the Best-Route strategy.  Second,
we can observe that the more next hops available for the ASF strategy the lower the loss rates, and a multi-path factor of 4 has nearly no loss for all the nodes.

The message overhead for link-state (LS) and HR is shown in Table~\ref{tab:ndn-testbed-overhead}.  In the no-failure case, HR has a slightly lower overhead
than that of link state and the difference between the various multi-path factors is very small.  This is not surprising as the probing algorithm has a fixed probing interval so each node generates the same number of probes for each route entry regardless of the number of next hops in the route entry, but the probes may cross different number of hops depending on the paths they travel, so the total number of probe Interests forwarded by all the nodes will vary slightly.  For an Internet-scale topology, however, there may be nodes with tens or hundreds of neighbors.  Using all next hops on these highly-connected nodes would be infeasible as each entry in the Route Cache will be very large and the statistics of the next hops will consume a significant amount of memory.  Moreover, next-hop selection will be very time-consuming.

Based on the above results,
we use \emph{a multi-path factor of 4 and a probing interval of 60 seconds for HR and ASF} in the remaining experiments.
With these parameters, HR with ASF has lower overhead than LS but is still effective in finding paths during failure and recovery periods.

\subsection{Message Overhead under Different Traffic Profiles}

We now use the 22-node topology and no-failure scenario to evaluate the message overhead under different traffic profiles.  More specifically, each node pings a percentage $\alpha$ of all the name prefixes in the network at 1 ping/second per name prefix.  The pinged prefixes are uniformly selected, so each prefix is pinged by $\alpha N$ nodes.  We vary $\alpha$ from 10\% to 100\%.

Since ASF probes every name prefix seen in recent Interests (we call them ``\textbf{active name prefixes}") at a constant rate, a node's overhead should increase with the number of active name prefixes at that node.
In contrast, NLSR's message overhead should be independent of the regular Interest/Data traffic as long as the latter does not congest the links, since NLSR's control messages are not triggered by traffic. 
\begin{figure}[tbh]
\centering
    \includegraphics[width=0.9\linewidth]{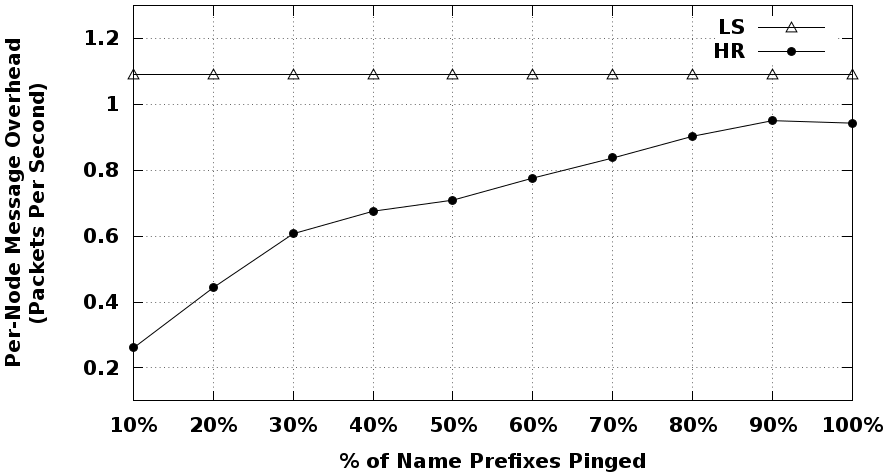}
\caption{Message Overhead in 22-Node Topology with Different Traffic Profiles (No Failures)}
\label{fig:testbed-traffic-scaling}
\end{figure}

Figure~\ref{fig:testbed-traffic-scaling} shows that, as expected, \emph{a smaller $\alpha$ is associated with lower HR/ASF probing overhead, while the LS overhead remains constant}.  In particular, when $\alpha$ is 10\%, the LS overhead is over four times that of HR/ASF. The results also show that the HR/ASF overhead does not increase linearly with $\alpha$.  The reason is that although each node pings only $\alpha N$ name prefixes, the number of active name prefixes at certain nodes can be much higher than $\alpha N$ if they are on the path of other nodes' Interests.  This phenomenon is most prominent when $\alpha$ is small and diminishes as $\alpha$ increases.

The implication of the above results is the following: HR/ASF is much more efficient than LS in an environment where users are not interested in all the name prefixes. We take user preference into account by having each node ping a percentage of name prefixes. but each node selects the pinged name prefixes uniformly, so every name prefix is pinged by the same number of nodes and produces the same amount of ping replies.  In practice, however, the data distribution among name prefixes is more likely to be highly skewed (e.g., 90\% of data is produced under 10\% of name prefixes), as shown in previous studies (\cite{Kim:2009,StabilityPopDest}). Therefore, our experiments likely over-estimated the number of active name prefixes and thus the probing overhead of HR/ASF in a real network should be even lower than in our experiments.

\subsection{Stress Test: Sequential Node Failure Scenario}

We continue to use the 22-node topology but introduce sequential failures of the 10 most connected nodes -- more specifically, after the first 60 seconds, a different node fails for 90 seconds every 180 seconds until all the 10 nodes have failed once.  Every node pings 10\% of the name prefixes, i.e., $\alpha = 10\%$, at 1 ping per second.

\begin{figure}[tbh]
\centering
    \includegraphics[width=0.9\linewidth]{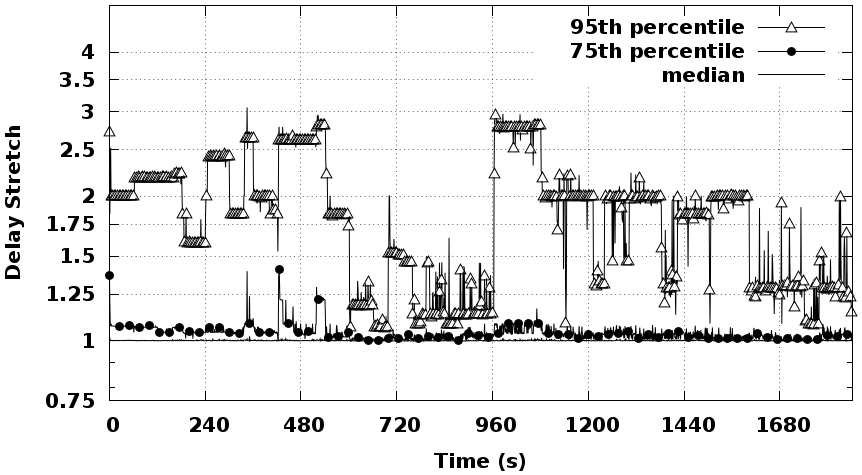}
\caption{Delay Stretch for HR in 22-Node Testbed Topology under Sequential Failure of 10 Most Connected Nodes and $\alpha = 10\%$ (vertical dash lines indicate when nodes fail and recover.)}
\label{fig:multiple-failure-stretch}
\end{figure}

\begin{figure}[tbh]
\centering
        \includegraphics[width=0.85\linewidth]{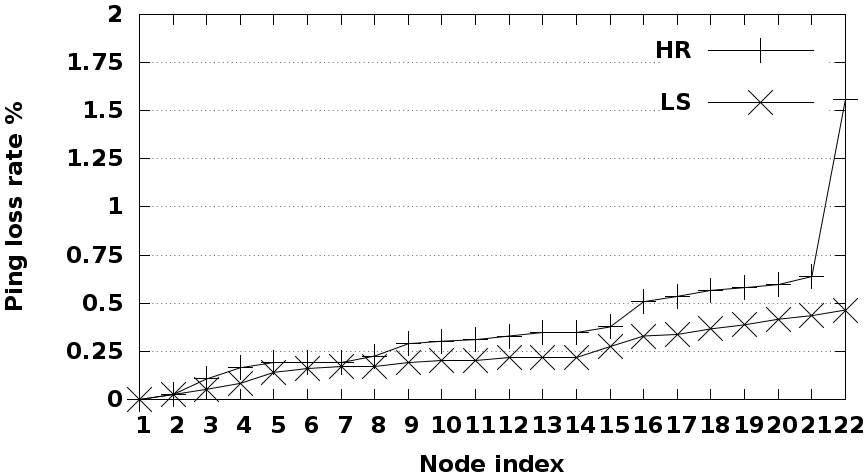}
    \caption{Loss Rate in 22-Node Topology under Sequential Node Failures (note that we exclude pings sent to the failed nodes from the loss rate calculation.)}
    \label{fig:multiple-failure-loss-rate}
\end{figure}

Figure~\ref{fig:multiple-failure-stretch} shows that, under a series of node failures, the median of the delay stretch remains near 1 for the entirety of the experiment.
The average values of the 75th-percentile and the 95th-percentile stretch are 1.04 and 1.82, respectively. 
The largest 75th-percentile stretch and 95th-percentile stretch are 1.41 and 3.07, respectively.
HR and LS have an average loss rate of 2.62\% and 2.48\%, respectively.
However, as Figure~\ref{fig:multiple-failure-loss-rate} shows, one of the nodes has a large difference in loss rate, which is under investigation.
Overall these results indicate that the packet delay under HR with ASF approximates that of LS.
Meanwhile, the HR/ASF approach has room for improvement in minimizing losses.  Note, however, that the node failures occurring every 180 seconds is an extreme scenario intended to stress our design, so one should not use the above loss rates to infer HR/ASF's performance in real networks.

\begin{figure}
\centering
    \begin{minipage}{0.4\textwidth}
        \centering
        \includegraphics[width=0.9\columnwidth]{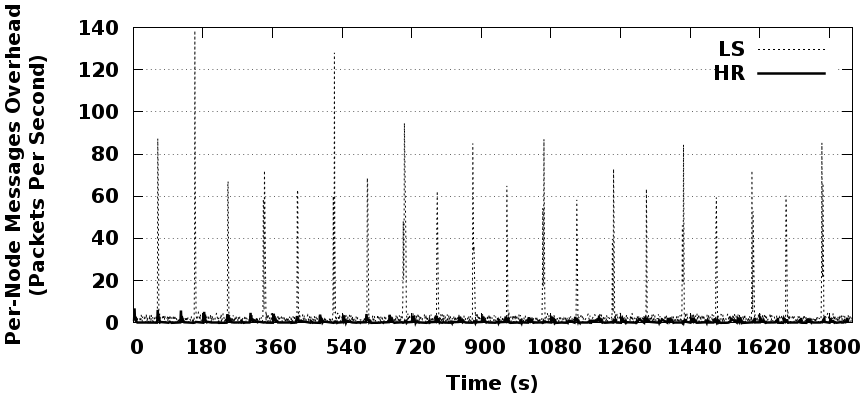}
        \label{fig:multiple-failure-overhead1}
        \vspace{-2ex}
    \end{minipage}
    \begin{minipage}{0.4\textwidth}
        \centering
       \includegraphics[width=0.9\columnwidth]{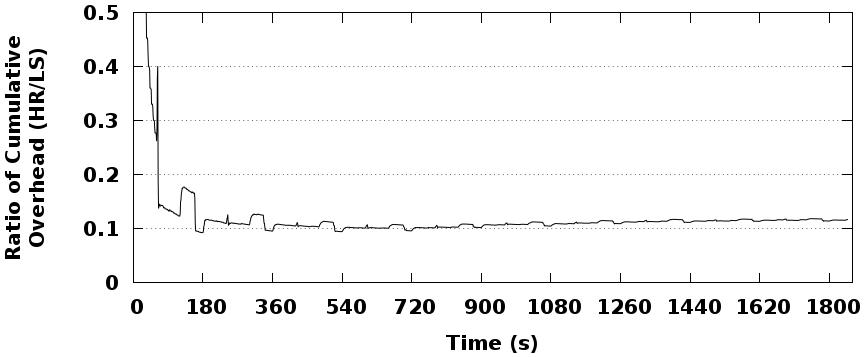}
       \label{fig:multiple-failure-overhead2}
    \end{minipage}
   \caption{Message Overhead in 22-Node Topology under Sequential Failure of 10 Most Connected Nodes and $\alpha = 10\%$}
    \label{fig:multiple-failure-overhead}
\end{figure}

As shown in Table~\ref{tab:ndn-testbed-overhead},
link state's overhead under sequential failures is 2.80 packets/second, which is 8.5 times the overhead of HR, 0.33 packets/second.  Note that $\alpha$ is 10\% in this experiment so HR's probing overhead is less than that when all the name prefixes are pinged.  Figure~\ref{fig:multiple-failure-overhead} shows the instantaneous overhead of HR and LS as well as the ratio between HR and LS's cumulative overhead over time.
For a short period at the beginning, HR has more overhead than LS due to probes being triggered by pings, but HR's overhead is quickly overtaken by LS's overhead. HR's overhead is nearly a 10th of LS's overhead for the remainder of the experiment.
LS's routing overhead also has instantaneous overhead spikes reaching as high as 138 packets per second when node failures and recoveries happen, while HR's probing overhead remains below 7 packets per second and only increases at probing intervals (60 seconds). These results confirm that \emph{HR/ASF's probing overhead is primarily determined by its probing interval and the number of active name prefixes in the network\footnote{Topological changes such as node failures and recoveries may change the paths probes take and affect the probing overhead accordingly, but path length differences are small in highly connected networks.}, while link-state routing's overhead increases as the network dynamics increase since it must send LSA Interests/Data as well as SYNC Interest/Data for each failure and recovery}.

\subsection{Scaling to Larger Topologies}

To study how HR's performance and overhead scale with the network size, we run experiments on five topologies ranging from 22 nodes to 99 nodes (Section~\ref{sec:method}), with sequential failure of the 10 most connected nodes in each experiment.   Each node pings 10\% of the name prefixes, e.g., 4 name prefixes in the 41-node topology and 10 name prefixes in the 99-node topology. This is a rather crude approximation of the ``network effect" -- the larger the network, the more resources a consumer can retrieve and the more people with which the consumer may want to communicate.

\begin{figure}
\centering
    \begin{minipage}{0.4\textwidth}
        \centering
        \includegraphics[width=\linewidth]{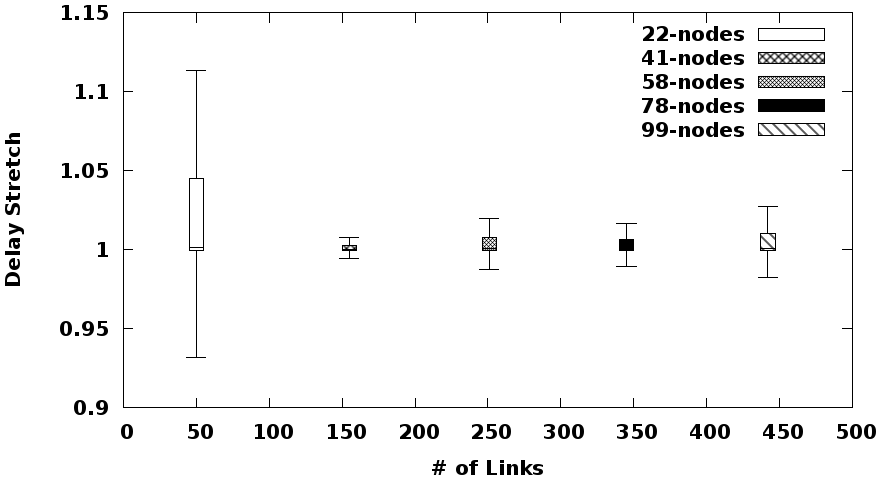}
        \mbox{\footnotesize a) Delay Stretch}
        \vspace{2ex}
    \end{minipage}
    \begin{minipage}{0.4\textwidth}
        \centering
        \vspace{-2ex}
        \includegraphics[width=\linewidth]{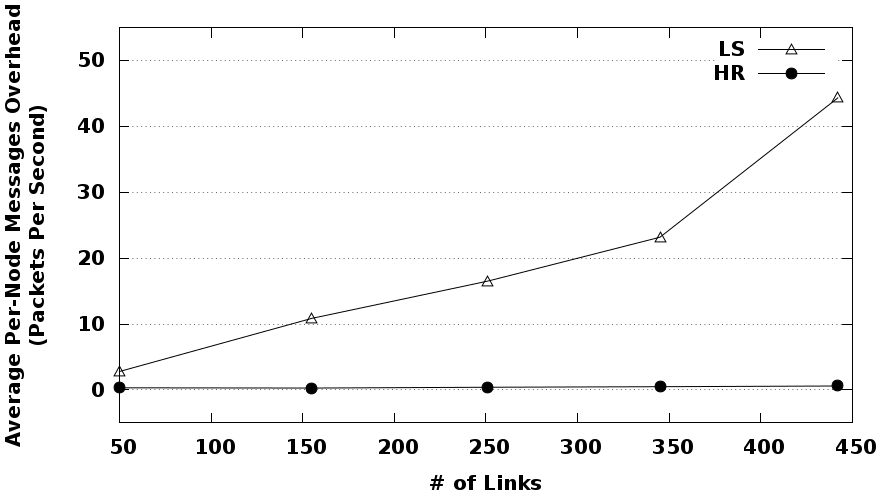}
        \mbox{\footnotesize b) Message Overhead}
        \vspace{2ex}
    \end{minipage}
    \vspace{-2ex}
   \caption{HR Performance as a Function of Increasing Network Size (Sequential Failure of 10 Most Connected Nodes and $\alpha = 10\%$)}
    \label{fig:large-topology-failure}
\end{figure}

Figure~\ref{fig:large-topology-failure}(a) shows the delay stretch as a function of the number of links in the topology\footnote{The two ends of a box represent the 25th and 75th percentile values and the whiskers extend the box by 1.5 times the interquartile range or to the most extreme data value in that range. The lower end of the delay stretch is sometimes below 1 because of small variations in RTT in different experiments. For example, a ping packet in the LS experiment may have experienced some queuing delay but the corresponding packet in the HR experiment may have had an un-congested path.}.
The four Internet-derived topologies have rather small delay stretches close to 1, much smaller than that for the 22-node testbed network, perhaps because they have richer connectivity as indicated by their higher average degree (from 7.56 to 8.93) compared to the testbed topology (4.55).  The richer the network connectivity, the smaller the difference between the best path and suboptimal paths and the easier it is for the forwarding plane to find alternative paths to get around loops and failures.

Figure~\ref{fig:large-topology-failure}(b) shows that the message overhead of HR is much lower than that of LS in all the topologies.
From the 22-node to the 99-node topology, HR's overhead grows very slowly from 0.33 packets/sec to 0.6 packets/sec and LS's overhead grows from 2.80 packets/sec to 44.32 packets/sec.
More specifically, the LS overhead first increases linearly until 345 links (78-nodes) and then increases sharply from 345 links to 442 links (99-nodes).
Our results suggest that HR scales much better than LS while achieving similar data delay as LS.

\com{
There are increases around the failure and recovery time, but overall the delay stretch is consistent with earlier results.  For the 41-node topology, the per-node message overhead is 4.506 and 0.28 pps for LS and HR, respectively.  For the 58-node topology, the per-node message overhead is 5.88 and 0.38 pps for LS and HR, respectively.  Even if we scale the HR overhead numbers by 10 (considering each node is pinging only 10\% of the nodes), they are still smaller than those of LS.
}

\section{Related Work} \label{sec:related}

Yi et al. (\cite{Forwarding-COMCOM,Cheng:RoleOfRouting}) have shown that NDN's adaptive forwarding plane can detect and recover from link failures quickly and can enable the use of routing protocols that were previously viewed as unsuitable for real networks. This is one of the reasons why HR may work well in NDN and help address its routing scalability problem (Section~\ref{sec:intro}).

We made the decision to develop our own ASF strategy, because currently proposed and implemented strategies, including the one proposed in Yi et al.'s work (\cite{Forwarding-COMCOM,Cheng:RoleOfRouting}), are either not available in NFD or do not sufficiently meet our requirements (Section~\ref{sec:strategy}).  Note that our objective for this work is not to develop the optimal strategy, but to get a baseline performance of HR with a basic forwarding strategy; further improvement on the strategy will lead to even better HR performance.  Our design is influenced by the following two forwarding strategies.  Chiocchetti et al.~\cite{strategy-INFORM} developed an ICN forwarding algorithm called INFORM which uses probing to learn the delay between nodes; this knowledge is then used to choose an interface to forward more effectively. INFORM uses RTTs piggybacked on Data to get an accurate estimate for Data retrieval.   Unlike INFORM, our design does not require changes to the Interest/Data format.  Qian et al.~\cite{probability-adaptive-forwarding} presented an adaptive forwarding strategy that uses an ant colony optimization based algorithm to probabilistically select interfaces for probing.  We also use past measurements to influence the probability of picking an interface for probing, but our algorithm is simpler and easier to analyze.  

To evaluate the feasibility of HR on the NDN testbed, the authors of~\cite{greedyforward} simulated different overlay topologies of the NDN hub nodes by connecting each node to its $m = 1, 2, 3$ hyperbolically closest neighboring nodes. For each value of $m$, they measured the success ratio and three types of stretch measures, with all possible removals of one link and one node. They found that although $m=2$ is enough for a 100\% success ratio on the NDN testbed, they recommend using $m=3$ for greater reliability.  Rather than making the overlay topology more resilient under HR, our work focuses on using smart forwarding strategy and multi-path forwarding to improve the packet delivery ratio.

\section{Conclusion}
\label{sec:concl}

We have evaluated some important scalability characteristics of HR in NDN.
HR is an attractive candidate to address NDN routing scalability issues since it does not require the precise full knowledge of the topological connectivity of highly dynamic networks.
Our evaluations shows that HR's forwarding plane overhead in such networks is much lower than NLSR's control overhead, while achieving similar packet delay and loss rates.
Most importantly, we have shown that as a function of the network size, HR's per-node message overhead is almost constant and close to zero, versus the usual polynomial growth experienced by traditional links-state routing.
Taken together, these results suggest that HR has immense potential to be an extremely scalable routing solution for NDN and possibly other networks.
Yet, several technical issues require further investigation.
These include mechanisms to compute and distribute hyperbolic coordinates in a decentralized manner, properly taking into account link delays.

\section{Acknowledgments}

This work was supported by NSF Grants 0964236, 1040036, 1039615, 1040868, 1344495, 1345142, 1345286, 1345318, and 1441828.

\end{document}